\documentclass[a4paper,fleqn]{cas-dc}

\usepackage[authoryear,longnamesfirst]{natbib}
\usepackage{enumitem}
\usepackage{graphicx}
\usepackage{multirow}
\usepackage{amsmath,amssymb,amsfonts}
\usepackage{amsthm}%
\usepackage{mathrsfs}%
\usepackage{xcolor}%
\usepackage{url}
\usepackage{textcomp}
\usepackage{multirow}
\usepackage{mathrsfs}
\usepackage{booktabs}
\usepackage{color}
\usepackage{bm}
\usepackage{float}
\usepackage{multirow}
\usepackage{pifont}
\newcommand{\cmark}{\ding{51}}%
\newcommand{\xmark}{\ding{55}}%

\usepackage{stfloats}

\def\tsc#1{\csdef{#1}{\textsc{\lowercase{#1}}\xspace}}
\tsc{WGM}
\tsc{QE}
\tsc{EP}
\tsc{PMS}
\tsc{BEC}
\tsc{DE}

\begin{document}
\let\WriteBookmarks\relax
\def\floatpagepagefraction{1}
\def\textpagefraction{.001}

\let\printorcid\relax 
 
\shorttitle{}


\title [mode = title]{SSH-Net: A Self-Supervised and Hybrid Network for Noisy Image Watermark Removal} 





\author[1]{Wenyang Liu}
\ead{wenyang001@e.ntu.edu.sg}
\affiliation[1]{
    organization={School of Electrical and Electronic Engineering, Nanyang Technology University},
    country={Singapore}}

\author[1]{Jianjun Gao}
\ead{gaoj0018@e.ntu.edu.sg}

\author[1]{Kim-Hui Yap}
\ead{ekhyap@ntu.edu.sg}

\cortext[cor1]{Corresponding author.}

\cormark[1]



\begin{abstract}
Visible watermark removal is challenging due to its inherent complexities and the noise carried within images.
Existing methods primarily rely on supervised learning approaches that require paired datasets of watermarked and watermark-free images, which are often impractical to obtain in real-world scenarios.
To address this challenge, we propose SSH-Net, a Self-Supervised and Hybrid Network specifically designed for noisy image watermark removal. SSH-Net synthesizes reference watermark-free images using the watermark distribution in a self-supervised manner and adopts a dual-network design to address the task. 
The upper network, focused on the simpler task of noise removal, employs a lightweight CNN-based architecture, while the lower network, designed to handle the more complex task of simultaneously removing watermarks and noise, incorporates Transformer blocks to model long-range dependencies and capture intricate image features.
To enhance the model’s effectiveness, a shared CNN-based feature encoder is introduced before dual networks to extract common features that both networks can leverage. 
Comprehensive experiments show that our proposed method surpasses state-of-the-art approaches in both performance and efficiency, demonstrating its effectiveness in noisy image watermark removal. 
Our code will be available at https://github.com/wenyang001/SSH-Net. 
\end{abstract}




\maketitle

\section{Introduction}

Social media platforms have become essential channels for sharing and distributing multimedia content, such as images and videos, making the security and robustness of digital media critical areas of research. Among various protection strategies, watermarking~\cite{singh2013survey, dekel2017effectiveness, hu2005algorithm} remains a widely used technique for copyright enforcement. These watermarks, typically in the form of text, numbers, or logos, are embedded into images to assert ownership and prevent unauthorized use. However, while watermarks provide effective copyright protection under certain conditions, they face challenges regarding robustness and resilience against unauthorized removal, especially as adversarial techniques advance. 
With the rise of AI-Generated Content (AIGC), these challenges have become more difficult, as automated tools can effortlessly manipulate or add watermarks, introducing new complexities in protecting and authenticating digital content.

In parallel with the development of watermarking, researchers have focused on watermark removal methods as an adversarial approach to evaluate and enhance these techniques. Early watermark removal methods typically relied on a composition model, where the watermarked image is decomposed into a watermark-free image and the watermark itself. For example, the Independent Component Analysis (ICA) algorithm proposed in~\cite{pei2006novel} attempts to separate these components to recover the original image.
Building on this, a probabilistic method~\cite{hsu2011new} models the relationships among the energies of the original image, the watermark, and the watermarked image to improve watermark removal performance. However, accurately estimating the watermark-free image is non-trivial and typically faces several challenges, such as separating overlapping patterns between the watermark and the original content without degrading image quality.

To address these challenges, many studies have treated the watermark removal task as an image-to-image translation problem. Advanced deep learning frameworks, including convolutional neural networks (CNNs) and generative adversarial networks (GANs), have been applied to improve watermark removal performance~\cite{dekel2017effectiveness, cheng2018large, huang2004attacking}. 
While these methods have achieved notable results, they typically rely on ground truth watermark-free images for training, limiting their applicability in real-world scenarios where such data may be scarce or unavailable. 
Inspired by recent advances in self-supervised learning for image restoration, researchers have begun to explore these techniques for the task of watermark removal, which similarly suffers from the lack of paired clean data in real-world settings. Self-supervised learning has shown remarkable success in tasks such as denoising~\cite{lehtinen2018noise2noise, krull2019noise2void} and dehazing~\cite{liang2022self} by learning directly from the corrupted image itself. by learning directly from the corrupted image itself. For instance, 
Noise2Noise~\cite{lehtinen2018noise2noise} demonstrated that networks can be trained using pairs of independently corrupted images without requiring clean targets, under the assumption of zero-mean noise. In contrast, Noise2Void~\cite{krull2019noise2void} introduced masking-based strategies that predict missing pixels from their spatial context, thereby eliminating the need for paired training data altogether. Inspired by such masking-based strategies, Liang~\textit{et al.}~\cite{liang2022self} applied a self-supervised approach to single image dehazing, incorporating haziness-guided masking to enable network training without relying on ground truth supervision.
These methods demonstrate the potential of self-supervised frameworks to generalize across restoration problems where clean data is difficult to obtain. Recently, this line of work has been extended to the watermark removal domain, where degradations are often more structured and localized. 
For example, Tian~\textit{et al.}~\cite{tian2024perceptive} proposed PSLNet, a state-of-the-art self-supervised learning network for watermark removal in noisy images. PSLNet adopts a parallel network architecture, where the upper network sequentially removes noise and watermarks, while the lower network handles both tasks simultaneously. Although this dual-network design demonstrates strong performance, its reliance on four identical CNN-based U-Net structures results in high computational costs and increased parameter complexity.
Furthermore, using the same architecture for sub-tasks with varying levels of complexity leads to inefficiencies, as the model lacks the flexibility to adapt to the specific needs of each task.

To address these challenges, we propose a Self-Supervised and Hybrid Network (SSH-Net) for noisy image watermark removal. SSH-Net adopts a dual-network design similar to~\cite{tian2024perceptive}.
However, rather than using identical architectures for both paths, the upper network employs a more efficient CNN architecture optimized for noise reduction, enabling it to operate with reduced computational cost. In contrast, the lower network, focused on the more complex task of simultaneously removing watermarks and noise, utilizes a deeper architecture that incorporates a sparse Transformer-based U-Net. This design allows the model to capture intricate features and model complex relationships more effectively, resulting in superior performance in challenging scenarios.
Furthermore, to ensure the model effectively leverages the complementary strengths of both networks, a shared feature encoder based on CNNs is introduced before the dual paths. The outputs from these paths are then fused using a gate mechanism, which dynamically balances their contributions. A mixed loss function is applied to each path, ensuring that both networks are optimally trained for their respective tasks, resulting in a final output with enhanced texture reconstruction. Experimental results demonstrate that our approach outperforms state-of-the-art methods in noisy image watermark removal while maintaining lower computational costs.
The contribution of this paper can be summarized as:

\begin{itemize}[itemsep=2pt,topsep=0pt,parsep=0pt]
    \item A self-supervised, hybrid dual-network approach combining CNNs and Transformers is proposed for noisy image watermark removal, eliminating the need for reference watermark-free images.
    \item The proposed method decomposes the task into two sub-tasks, each utilizing specially designed networks with varying depths and architectures to optimize the learning process. Additionally, a sparse Transformer U-Net is introduced into the network, enabling it to achieve state-of-the-art performance.
    \item A shared feature encoder is introduced before the dual networks to extract common features that both networks can leverage, and a gate mechanism is applied afterward to dynamically balance their contributions, thereby improving the overall performance.
\end{itemize}


The remainder of this paper is organized as follows: Section~\ref{s:RW} reviews related works, Section~\ref{s:M} presents details of our proposed method, Section~\ref{s:E} provides extensive experiments, performance analysis, and ablation studies, and Section~\ref{s:C} concludes the paper.

\section{Related Work}
\label{s:RW}
In this section, we first provide a comprehensive review of image watermark removal techniques. We then introduce the leading technique, Transformers, in image restoration, which is closely related to noisy image watermark removal.
\subsection{Early Studies in Image Watermark Removal}
Braudaway \textit{et al.}~\cite{braudaway1997protecting} were pioneers in the use of visible watermarks in digital images for ownership identification. In contrast to watermark embedding, numerous studies~\cite{cheng2018large, dekel2017effectiveness, huang2004attacking, pei2006novel, xu2017automatic} have focused on the removal of watermarks from watermarked images to restore the original image content. Early approaches primarily relied on hand-crafted features for watermark extraction. For example, Pei \textit{et al.}~\cite{pei2006novel} applied Independent Component Analysis (ICA) to separate the source image from the watermarked image. Dekel \textit{et al.}~\cite{dekel2017effectiveness} presented a multi-image matting algorithm to automatically estimate the alpha matte and remove visible watermarks. Despite these efforts, accurately estimating the watermark-free image and restoring the original content without degrading image quality remains challenging, especially with the rise of AI-Generated Content (AIGC), which can easily manipulate or add watermarks.

\subsection{Recent Studies in Image Watermark Removal}
Recently, some studies have tried to view the watermark removal task as an image-to-image translation problem, leveraging data-driven neural network methods to enhance performance and achieve more reliable results. For instance, a CNN-based network is trained on pairs of watermarked and watermark-free images to remove watermarks~\cite{geng2020real}, allowing the model to map watermarked images to a watermark-free domain. To enhance the robustness and precision of watermark removal, Cheng \textit{et al.}~\cite{cheng2018large} combined a watermark detector with an image translation model. This integrated approach first utilizes a watermark detector, built on well-established deep learning-based general object detection methods~\cite{zhao2019object}, to precisely locate watermarks in the original images. Once detected, the image translation model reconstructs the watermark-free images with minimal artifact introduction. 
More recently, generative adversarial networks (GANs) have been demonstrated to significantly improve watermark removal techniques, especially in handling complex and diverse watermarking scenarios.
Li \textit{et al.}~\cite{li2019towards} leveraged a patch-based discriminator, conditioned on watermarked images, to generate watermark-free reconstructions with photo-realistic details. Chen \textit{et al.}~\cite{chen2019leveraging} developed a carefully designed fine-tuning framework that effectively removes watermarks, even with limited labeled data.

\subsection{Transformer in Image Restoration}
Watermark removal follow a similar procedure to various image restoration tasks, as both involve reconstructing a degraded image to enhance its visual quality while addressing specific artifacts~\cite{liu2023bitstream, liang2021swinir}. Specifically, watermark removal aims to restore the original content of an image by removing watermarks, much like other restoration tasks that focus on removing artifacts such as haze~\cite{valanarasu2022transweather, song2023vision}, rain~\cite{chen2023learning, xiao2022image}, or shadows~\cite{chang2023tsrformer, wan2024crformer}. 
With the rise of vision Transformers~\cite{dosovitskiy2020image, ByteNet, liu2021swin}, their ability to model long-range dependencies and capture global contextual information has led to their application in various image restoration tasks. Notably, SwinIR~\cite{liang2021swinir} and Restormer~\cite{zamir2022restormer} are two prominent Transformer-based models that have been proposed to address image restoration challenges.
SwinIR employed a hierarchical design using Swin Transformer blocks, which progressively capture global dependencies in images. Restormer, on the other hand,  adopted a U-Net structure and integrated Transformer blocks to enhance feature extraction and reconstruction capabilities. Song \textit{et al.}~\cite{song2023vision} proposed a similar structure to SwinIR with the modified normalization layer and activation functions to improve image dehazing results. In contrast, Valanarasu \textit{et al.}~\cite{valanarasu2022transweather} employed a DETR-like framework to handle multiple weather conditions simultaneously. Chen \textit{et al.}~\cite{chen2023learning} proposed a efficient Transformer-based structure for image deraining to help to generate accurate details. To address shadow removal, Chang \textit{et al.}~\cite{chang2023tsrformer} proposed a two-stage architecture, TSRFormer, which integrates Transformer with a content refinement process. 

In this work, we tackle a more complex watermark removal task that involves significant noise, and propose a hybrid structure by combining both CNN and Transformer.

\begin{figure*}
\centering
\includegraphics[width=0.99\linewidth]{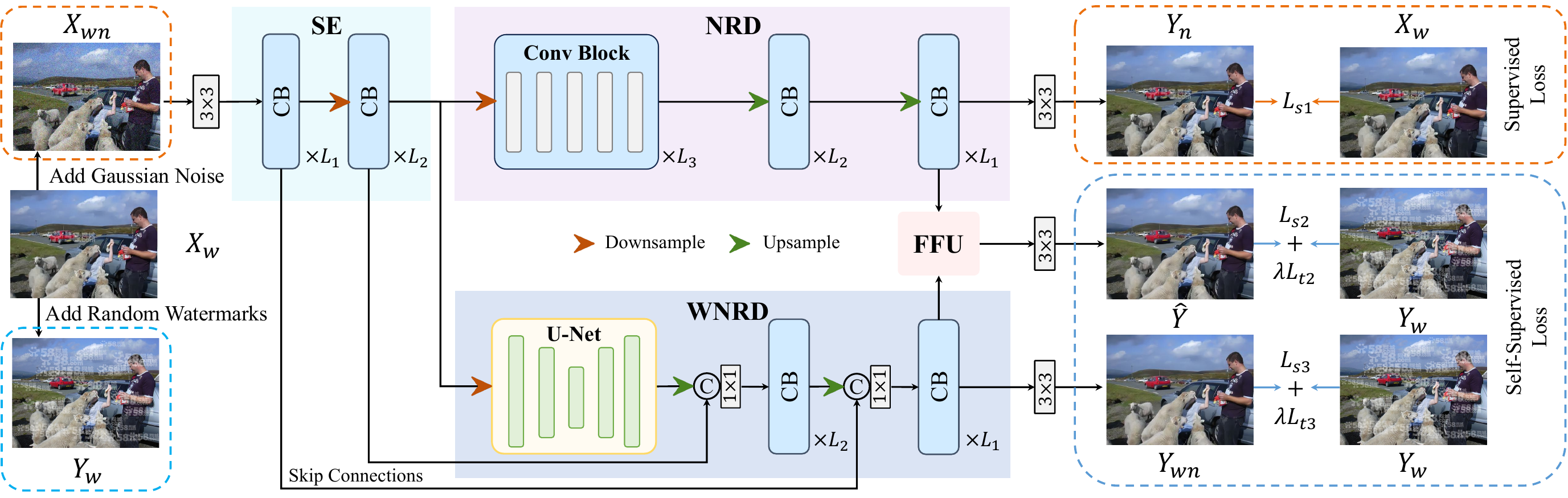}
    \caption{The overall architecture of the proposed Self-Supervised and Hybrid Network (SSH-Net), which mainly contains a shared encoder, a noise removal decoder (NRD), a watermark and noise removal decoder (WNRD), and a feature fusion unit (FFU). SSH-Net leveraged the supervised and self-supervised technique to generate groud-truth samples $(X_{\text{w}}, Y_{\text{w}})$ to optimize the model. }
    \label{fig:overview}
    \vspace{-0.25in}
\end{figure*}

\section{Proposed Method}
\label{s:M}
In this section, we first provide the preliminary knowledge of self-supervised learning and the formulation of the noisy image watermark removal problem. Next, we present the overall architecture of the proposed SSH-Net and describe its key components.

\subsection{Preliminary and Problem Formulation}
A typical watermark removal task aims to restore the clean image $Y$ from a watermarked image $X_{\text{w}}$ generated by blending a watermark $W$ into $Y$, represented as:

\vspace{-0.2in}
\begin{equation}
\label{eq:water}
    X_{\text{w}}(p) = \alpha(p)W(p) + (1 - \alpha(p))Y(p),
\end{equation}

\noindent where $p = (i, j)$ denotes the pixel location in the image, and $\alpha(p)$ is a spatially varying opacity that controls the visibility of the watermark.
Existing watermark removal methods mainly address it in a supervised manner. The corresponding objective is to train a regression model, using the paired training data, by minimizing the expectation of loss:


\vspace{-0.2in}
\begin{equation}
\arg\min_{\theta}  E_{(X_{\text{w}}, Y)}\{L(f_\theta(X), Y)\}, 
\end{equation}


\noindent where $f_\theta(\cdot)$ denotes the model with parameters $\theta$, $L$ is a loss function (e.g., L2 or L1 loss), and $E_{(X_{\text{w}}, Y)}\{\cdot\}$ represents the expectation over the paired data. 
However, acquiring paired data can be difficult. A common solution is to break down the whole training process into two different optimization phases, as follows:


\vspace{-0.2in}
\begin{equation}
\label{eq:two}
\arg\min_{\theta}  E_{X_\text{w}}\{ E_{Y|X_\text{w}} \{ L(f_\theta(X_\text{w}), Y)\} \}, 
\end{equation}

\noindent where $E_{X_\text{w}}\{\cdot\}$ and $E_{Y|X_\text{w}}\{\cdot\}$ denote the expectation and conditional expectation, respectively.
But still, the reference watermark-free images $Y$ are not easy to obtain, which limits the applicability of such approaches in real-world scenarios.

To address this issue, some recent studies~\cite{tian2024perceptive, tian2024self} have leveraged self-supervised technique~\cite{lehtinen2018noise2noise} to generate synthesized images that can serve as substitutes for the watermark-free samples. Specifically, the synthesized reference image $Y_\text{w}$ is generated by randomly adding additional watermarks into the watermarked image $X_\text{w}$. This ensures that both the input image and the synthesized output image are drawn from the same watermark corruption distribution, approximating the condition $E\{Y_\text{w}|X_\text{w}\} = Y$. In other words, $Y_\text{w}$ serves as an unbiased estimator of the clean image $Y$, conditional on the watermarked image $X_\text{w}$. This unbiasedness guarantees that $Y_\text{w}$ accurately represents the real underlying ground-truth $Y$ on average, even though individual samples contain various corruption. Consequently, Eq.~\ref{eq:two} can be formulated as an empirical risk minimization task:

\vspace{-0.2in}
\begin{equation}
\label{eq:emp}
\arg\min_{\theta}  \frac{1}{N} \sum_{i=1}^N L(f_\theta(X_\text{w}^i), Y_\text{w}^i),
\end{equation}

\noindent where $X_\text{w}^i$ represents an input watermarked image, $Y_\text{w}^i$ represents a synthesized ground-truth image, and $N$ denotes the number of samples in the dataset. Therefore, minimizing the discrepancy between the prediction $f_{\theta}(X_\textbf{w})$ and $Y_\text{w}$ leads the model to approximate the real ground-truth Y, even without direct supervision.

As noise is often introduced during the transmission or storage of watermarked images, it is crucial to develop a robust model that can adeptly handle both watermark and noise degradations. To address this challenge, we focus on the task of noisy image watermark removal in this paper. The noisy watermarked image is constructed as $X_{\text{wn}} = X_{\text{w}} + K$, where $K$ denotes the Gaussian noise added to the original watermarked image $X_{\text{w}}$. The objective of this task is to recover the clean image $Y$ from the watermarked noisy image $X_{\text{wn}}$ using synthesized data pairs $(X_{\text{w}}, Y_{\text{w}})$ during training, while relying solely $X_{\text{wn}}$ during inference. 


\subsection{Overall Structure}
The overall architecture of our proposed SSH-Net is shown in Fig.~\ref{fig:overview}, consisting of four main components: a shared encoder (SE), a lower watermark and noise removal decoder (WNRD), an upper noise removal decoder (NRD), and a feature fusion unit (FFU). Given that noise removal is a relatively simpler task and benefits from supervised learning, we design the NRD decoder using pure convolutional layers for fast and stable convergence. In contrast, WNRD handles both structured watermark and noise removal, which is inherently more complex. Therefore, we enhance WNRD with additional Transformer-based modules, including sparse attention layers, to better capture long-range dependencies and structured patterns in watermark regions. Despite the decoder differences, both branches share a common encoder, ensuring efficient feature reuse and consistent representation across the two tasks.

\textbf{\textit{Shared Encoder.}} To be specific, the watermarked noisy image input $X_{\text{wn}} \in \mathbb{R}^{H \times W \times 3}$ is first sent to a 3$\times$3 convolution layer to extract shallow feature $F_0 \in \mathbb{R}^{H \times W \times C}$, where $H, W, C$ denote the feature map's height, width, and channel dimensions, respectively. Subsequently, these shallow features $F_0$ are passed through the SE, which is composed of two Convolutional Blocks (CBs), specifically NAFBlocks~\cite{chen2022simple} (Nonlinear Activation Free Block), each followed by a downsampling layer, to progressively extract the shared features $F_{\text{se}} \in \mathbb{R}^{\frac{H}{4} \times \frac{W}{4} \times C}$:

\vspace{-0.2in}
\begin{equation}
F_{\text{se}} = H_\text{SE}(\text{Conv}(X_{\text{wn}})),
\end{equation}

\noindent where $H_\text{SE}(\cdot)$ represents the SE module, and Conv denotes a convolution layer. 

\textbf{\textit{Dual Decoders.}} The shared features $F_{\text{se}}$ are then split into two branches: one directed to the NRD to generate the feature $F_\text{n} \in \mathbb{R}^{H \times W \times C}$ for noise removal only, and the other directed towards the WNRD to produce the feature $F_{\text{wn}} \in \mathbb{R}^{H \times W \times C}$, targeting the removal of both noises and watermarks.


\vspace{-0.2in}
\begin{equation}
F_{\text{wn}} = H_\text{WNRD}(F_{\text{se}}) \ ; \ F_{\text{n}} = H_\text{NRD}(F_{\text{se}}),
\end{equation}


\noindent where $H_\text{NRD}(\cdot)$ and $H_\text{WNRD}(\cdot)$ represent the NRD and WNRD module, respectively. 

\begin{figure*}
\centering
\includegraphics[width=0.99\linewidth]{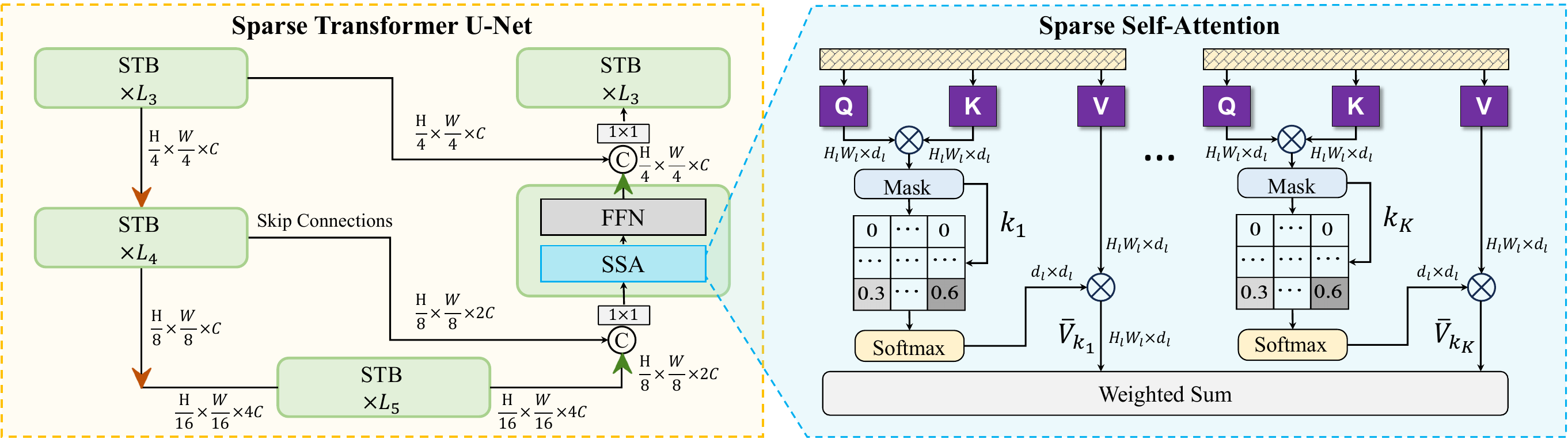}
    \caption{The overall architecture of the proposed Sparse Transformer U-Net, which follows a 3-level Transformer U-Net design. Each level comprises multiple Sparse Transformer Blocks (STBs) to process features at different scales. Each STB includes a Sparse Self-Attention (SSA) mechanism and a Fully-Connected Network (FFN).}
    \label{fig:block}
    \vspace{-0.25in}
\end{figure*}

\textbf{\textit{Feature Fusion Unit.}} Finally, these features are fused in FFU to generate the fused feature $F_{\text{fuse}} \in \mathbb{R}^{H \times W \times C}$. This process leverages $F_{\text{wn}}$ to create a gating signal that modulates $F_\text{n}$, which is then added to $F_{\text{wn}}$, followed by a NAFBlock to enhance the feature representation as:

\vspace{-0.2in}
\begin{align}
F_{\text{fuse}} &= H_\text{FFU} (F_{\text{wn}}, F_{\text{n}})\\ 
         &= \text{NAFBlock}(F_{\text{wn}} +  \text{Gating}(F_{\text{wn}}) \odot F_{\text{n}}) , 
\end{align}

\noindent where Gating$(\cdot)$ represents the gating mechanism, which consists of several convolution layers, and $\odot$ is the element-wise multiplication operator. The applied gating mechanism allows the model to adaptively select the most relevant features from $F_{\text{n}}$ to strengthen the representation. Based on the extracted features, we utilize three 3$\times$3 convolutional layers generate two outputs from the dual networks: $Y_{\text{wn}} = \text{Conv}(F_{\text{wn}})$ (watermark and noise removal output) and $Y_{\text{n}} = \text{Conv}(F_{\text{n}}) $ (noise removal output), as well as the final reconstructed clean output $\hat{Y} = \text{Conv}(F_{\text{fuse}}) + X_{\text{wn}}$. The intermediate outputs $Y_{\text{wn}}$ and $Y_{\text{n}}$ exclude residual connections to maintain diversity in feature representation. In contrast, the final output $\hat{Y}$ incorporates a residual connection with the initial input $X_{\text{wn}}$ to enhance reconstruction fidelity.




\subsection{Noise Removal Decoder}
The noise removal decoder (NRD), with a structure similar to the SE, employs a series of CBs aimed at effectively recovering the clean image $Y_{\text{n}}$ from the input features $F_{\text{se}}$. 
This process can be viewed as an auxiliary task that enhances the shared encoder's ability to focus on both watermark and noise patterns in the early part of the network. The decoder, together with the shared encoder, forms a classic 3-level convolutional U-Net architecture. Each level is specifically designed to process features at varying scales, with the middle level acting as a bottleneck to handle the features with smallest size in the network.


\textbf{\textit{Convolution Block.}} The core component of this decoder in each level lies in the convolution block, which is instantiated with the NAFBlock~\cite{chen2022simple}. The NAFBlock is a simplified yet efficient block designed to enhance computational efficiency by removing common nonlinear activation functions like Sigmoid, ReLU, and GELU. nstead, it utilizes convolution and element-wise multiplication operations to maintain performance while reducing complexity. Specifically, the NAFBlock adopts a Transformer-like structure but relies entirely on convolutional operations, such as Simplified Channel Attention (SCA), rather than self-attention mechanisms, to act as the \textit{token mixer}~\cite{yu2022metaformer, liang2021swinir}. Given the input feature map $F_{l-1}$ at the $l-1$-th NAFBlck, the whole process can be formulated as:



\vspace{-0.2in}
\begin{gather}
    F'_{l} = F_{l-1} + \text{Conv(SCA(SG(Conv(LN(}F_{l-1}\text{)))))}, \\
    F_{l} = F'_{l} + \text{Conv(SG(Conv(LN(}F'_{l}\text{))))},
\end{gather}

\noindent where $F'_{l}$ and $F_{l}$ denote the intermediate output feature map from the \textit{token mixer} stage and the final output feature map, respectively, SCA represents the Simplified Channel Attention, SG indicates the Simple Gate, and LN refers to the layer normalization. More details can be referred to~\cite{chen2022simple}.



\subsection{Watermark and Noise Removal Decoder}
The watermark and noise removal decoder (WNRD) serves as the core component for eliminating embedded watermarks while simultaneously reducing noise from the input features $F_{\text{se}}$. Together with the SE, it constitutes a 5-level hybrid U-Net architecture. The first two levels use convolutional blocks to extract features from the input image, while the remaining three levels employ Transformer blocks to process these features with downscaled size, capturing long-range dependencies. This hierarchical structure allows the model to capture both local and global dependencies while maintaining computational efficiency. 
Additionally, skip connections are employed at each level of the WNRD to merge features from the SE, ensuring efficient information flow across the various stages of the network.


\textbf{\textit{Sparse Transformer U-Net.}} 
The core component in WNRN is the proposed Sparse Transformer U-Net, placed centrally to handle downscaled features, aiming to enhance the model's representational capacity using Transformers. As illustrated in Fig.~\ref{fig:block}, it adopts a 3-level U-Net design, where each level comprises multiple Sparse Transformer (ST) blocks. Starting with the input features $F_{\text{se}}$, with spatial dimensions of $\frac{H}{4} \times \frac{W}{4}$, the first two levels in this U-Net progressively downscale and process the feature map using ST blocks until reaching the middle level, where the spatial dimensions are reduced to $\frac{H}{16} \times \frac{W}{16}$. The subsequent two levels then progressively upscale and process the feature map using ST blocks, ultimately restoring the spatial dimensions to $\frac{H}{4} \times \frac{W}{4}$, matching the input resolution. This component enhances the model’s ability to restore structured degradations. Together with the convolutional NRD branch, it forms a complementary design that effectively captures sparsely distributed watermark patterns, contributing significantly to the overall performance of SSH-Net when dealing with both noise and structured watermark artifacts.




\textbf{\textit{Sparse Transformer Block.}} Standard Transformers~\cite{liang2021swinir} typically face exponentially increasing computational complexity as the spatial dimensions of feature maps grow. Consequently, applying standard Transformers to most image restoration tasks, which often involve high-resolution inputs, is computationally inefficient. To address this limitation, we follow the design of Transformer proposed in~\cite{zamir2022restormer}, which introduces the multi Dconv head transposed attention (MDTA) layer. This layer applies self-attention across the channel dimension instead, reducing computational complexity from quadratic to linear. Furthermore, considering that watermark patterns are often localized and may not require attention across the entire spatial domain, we enhance the Sparse Transformer (ST) block by incorporating a sparse attention mechanism, i.e., top-$k$ sparse attention. Given the input feature map $F_{l-1}$ at the $l-1$-th ST block, the whole process can be formulated as:



\vspace{-0.2in}
\begin{gather}
    F'_{l} = F_{l-1} + \text{SSA(LN(}F_{l-1}\text{))}, \\
    F_{l} = F'_{l} + \text{FFN(LN(}F'_{l}\text{))},
\end{gather}

\noindent where SSA represents the sparse self-attention mechanism, which selectively focuses on relevant features based on the calculated attention scores, while FFN denotes the commonly used feed-forward network~\cite{zamir2022restormer}.

\textbf{\textit{Sparse Self-Attention.}} Sparse Self-Attention (SSA) is designed to selectively focus on specific elements within the input sequence, enabling a better contextual understanding of sparsely localized watermark patterns. Formally, as illustrated in Fig.~\ref{fig:block}, given a query $Q$, key $K$ and value $V$ with dimensions $H_lW_l \times d_l$ (where $H_l$ and $W_l$ represent the spatial dimensions, and $d_l$ is the channel dimensionality at the $l$-th ST block), $Q$ and $K$ are first reshaped and used to generate the attention map $M \in d_l \times d_l$. 
To enhance efficiency, an adaptive selection strategy is applied to mask out irrelevant elements, guided by a sequence of top-$k$ values, i.e., $(k_1, k_2, ..., k_K)$. For each $k_i$, the sparse attention is computed as follows:



\vspace{-0.2in}
\begin{align}
    \bar{V}_{k_i} &= \text{SparseAttention}(Q, K, V, \mathcal{S}_{k_i}) \\
    &= \text{Softmax} \left( \mathcal{S}_{k_i} \left( \frac{Q K^\top}{\lambda} \right) \right) V, 
\end{align}



\noindent where $\mathcal{S}_{k_i}(\cdot)$ is the learnable sparse selection operator, which retains the most critical elements based on the selected top-$k_i$ ranking and $\lambda$ is the scale factor, defined as:

\vspace{-0.2in}
\begin{equation}
    \left[\mathcal{S}_{k_{i}}\left(\frac{Q K^\top}{\lambda} \right) \right]_{ij} =
    \begin{cases}
        M_{ij}, & \text{if } M_{ij} \in \text{top-}k \ (\text{row } j), \\
        0, & \text{otherwise},
    \end{cases}
    \label{eq:topk_operator}
\end{equation}

\noindent where $M_{ij}$ represents the value of transposed attention map at position $(i,j)$. The final output $V_{out}$ is obtained by aggregating the results of all $\bar{V}_{k_i}$ across different $k_i$ as:

\vspace{-0.2in}
\begin{equation}
    \bar{V}_{out} = \frac{1}{K}\sum_{i=1}^K \bar{V}_{k_i}.
    \label{eq:head_output}
\end{equation}

\noindent The resulting output is further processed through layer normalization and subsequent feed-forward network layers to enhance feature representation and refinement.

\subsection{Loss Function}
As Fig.~\ref{fig:overview} shows, the proposed SSH-Net generates three images: $Y_\text{n}$ (image from the upper noise removal decoder), $Y_\text{wm}$ (image from the lower noise and watermark removal decoder), and $\hat{Y}$ (image from the final fused output). During the training stage, the data pairs $(X_\text{w}, Y_\text{w})$ are used as ground truth for both supervision and self-supervision, ensuring the model is effectively optimized. 
Following~\cite{tian2024perceptive}, a mixed loss function is employed, combining a structural loss $L_s$ (based on the L1 loss), and a texture loss $L_t$ (derived from a perceptual VGG network), expressed as:

\vspace{-0.2in}
\begin{equation}
    L = L_s + \alpha L_t, 
\end{equation}

\noindent where $\alpha$ is a hyperparameter that balances the contributions of the structure and texture losses. $L_s$ ensures pixel-level fidelity, helping the model preserve edges, object boundaries, and spatial structure during restoration. In contrast, $L_t$ captures perceptual similarity beyond raw pixel differences, guiding the model to recover visually realistic textures and suppress artifacts.


\textbf{\textit{Structural Loss.}} The structural loss $L_s$ ensures pixel-level fidelity by measuring the L1 difference. Specifically, $L_s$ consists of three parts: 

\vspace{-0.2in}
\begin{align}
L_s &= L_{s1} + L_{s2} + L_{s3} \\
 &= \frac{1}{N} \sum_{i=1}^N \Big( \left| Y_\text{n}^i - X_\text{w}^i \right| + \left| Y_{\text{wm}}^i - Y_\text{w}^i \right| + \left| \hat{Y}^i - Y_\text{w}^i \right| \Big),
\end{align}

\noindent where $L_{s1}$ is applied to the NRD output, $L_{s2}$ is applied to the NWRD output, and $L_{s3}$ is applied to the final FFU output. 

\textbf{\textit{Texture Loss.}} The texture loss $L_t$ evaluates high-level features extracted from a pretrained VGG network, guiding the model to preserve perceptual texture quality. Specifically, $L_{t}$ consists of two parts:

\vspace{-0.2in}
\begin{align}
L_t &= L_{t1} + L_{t2} \\
    &= \frac{1}{N} \sum_{i=1}^N  \left|\text{VGG}(Y_{\text{wm}}^i) - \text{VGG}(Y_\text{w}^i) \right| \nonumber \\
    &+  \frac{1}{N} \sum_{i=1}^N \left|\text{VGG}(\hat{Y}^i) - \text{VGG}(Y_\text{w}^i) \right|,
\end{align}


\noindent where $L_{t1}$ evaluates the perceptual differences between the output image $Y_{\text{wm}}$ and the synthesized ground-truth image $Y_\text{w}$, and $L_{t2}$ evaluates the perceptual differences between the reconstructed image $\hat{Y}$ from FFU and the synthesized ground-truth image $Y_\text{w}$.


\section{Experiment}
In this section, we first describe the experimental datasets and evaluation metrics. Next, we outline the experimental settings and compare our approach with state-of-the-art methods. Finally, we conduct ablation studies to assess the effectiveness of the proposed components.

\begin{table}[t]
\caption{PSNR, SSIM, and LPIPS comparison for noise levels of 0, 15, 25, and 50 with watermark transparency of 0.3.}
\label{table:1}
\resizebox{1.0\linewidth}{!}{
\begin{tabular}{ccccccc}
\hline
Methods  & PSNR$\uparrow$   & SSIM$\uparrow$   & LPIPS$\downarrow$  &  PSNR$\uparrow$   & SSIM$\uparrow$   & LPIPS$\downarrow$  \\ \hline
Noise levels           & \multicolumn{3}{c}{$\sigma$ = 0}  & \multicolumn{3}{c}{$\sigma$ = 15} \\ \hline
DnCNN         & 31.27  & 0.9482 & 0.0211 & 30.44  & 0.8833 & 0.1455 \\
FFDNet        & 28.82  & 0.8904 & 0.1019 & 29.03  & 0.8570 & 0.1755 \\
IRCNN         & 32.21  & 0.9824 & 0.0211 & 29.57  & 0.8734 & 0.1591 \\
FastDerainNet & 34.44  & 0.9807 & 0.0145 & 29.14  & 0.8550 & 0.1582 \\
DRDNet        & 31.97  & 0.9745 & 0.0305 & 27.24  & 0.8585 & 0.1706 \\
PSLNet  & \textcolor{blue}{42.16}  & \textcolor{blue}{0.9932} & \textcolor{blue}{0.0043} & \textcolor{blue}{32.07}  & \textcolor{blue}{0.8972} & \textcolor{blue}{0.1320}  \\ 
SSH-Net (Ours)  & \textcolor{red}{49.38}  & \textcolor{red}{0.9986} & \textcolor{red}{0.0017} & \textcolor{red}{32.24}  & \textcolor{red}{0.9026} & \textcolor{red}{0.1210} \\ \hline
Noise levels           & \multicolumn{3}{c}{$\sigma$ = 25} & \multicolumn{3}{c}{$\sigma$ = 50} \\ \hline
DnCNN         & 28.81  & 0.8231 & 0.2163 & 25.64  & 0.6934 & 0.3406 \\
FFDNet        & 26.84  & 0.7888 & 0.2509 & 25.17  & 0.6959 & 0.3537 \\
IRCNN         & 27.67  & 0.8008 & 0.2406 & 24.91  & 0.6795 & 0.3642 \\
FastDerainNet & 26.25  & 0.7799 & 0.2364 & 24.85  & 0.6821 & 0.3430 \\
DRDNet        & 26.53  & 0.8104 & 0.2280 & 25.83  & 0.7261 & 0.3090 \\
PSLNet         & \textcolor{blue}{29.82}  & \textcolor{blue}{0.8434} & \textcolor{blue}{0.1959} & \textcolor{blue}{26.90}  & \textcolor{blue}{0.7499} & \textcolor{blue}{0.2992} \\
SSH-Net (Ours)     & \textcolor{red}{29.86}  & \textcolor{red}{0.8533} & \textcolor{red}{0.1782} & \textcolor{red}{26.92}  & \textcolor{red}{0.7532} & \textcolor{red}{0.2876} \\
\hline
\end{tabular}
}
\vspace{-0.2in}
\end{table}

\begin{table}[t]
\caption{PSNR, SSIM, and LPIPS comparison for different methods with noise level of 25 with blind watermark transparency of 0.3, 0.5, 0.7, and 1.0.}
\label{table:2}
\resizebox{1.0\linewidth}{!}{
\begin{tabular}{ccccccc}
\hline
Methods            & PSNR$\uparrow$   & SSIM$\uparrow$   & LPIPS$\downarrow$  &  PSNR$\uparrow$   & SSIM$\uparrow$   & LPIPS$\downarrow$     \\ \hline
Transparency          & \multicolumn{3}{c}{Alpha = 0.3} & \multicolumn{3}{c}{Alpha = 0.5} \\ \hline
DnCNN         & 27.01   & 0.8008   & 0.2333   & 26.95   & 0.8006   &  0.2339   \\
FFDNet        & 25.34   & 0.7630   & 0.2781   & 23.97   & 0.7570   & 0.2834   \\
IRCNN         & 26.30   & 0.8084  & 0.2392   & 26.13   & 0.8064   & 0.2415   \\
FastDerainNet & 25.89   & 0.7724   & 0.2456   & 25.90   & 0.7720   & 0.2463   \\
DRDNet        & 24.01   & 0.7720   & 0.2630   & 24.51   & 0.7731   & 0.2625   \\
PSLNet         & \textcolor{blue}{28.43}   & \textcolor{blue}{0.8335}   & \textcolor{blue}{0.2078}   & \textcolor{blue}{28.01}   & \textcolor{blue}{0.8311}   & \textcolor{blue}{0.2104}   \\
SSH-Net (Ours)     & \textcolor{red}{28.97}  & \textcolor{red}{0.8410} & \textcolor{red}{0.1919} & \textcolor{red}{29.02}  & \textcolor{red}{0.8403} & \textcolor{red}{0.1928} \\

\hline
Transparency          & \multicolumn{3}{c}{Alpha = 0.7} & \multicolumn{3}{c}{Alpha = 1.0} \\ \hline
DnCNN         & 27.65   & 0.8021   & 0.2332   & 21.40   & 0.7878   & 0.2457   \\
FFDNet        & 22.86   & 0.7545   & 0.2856   & 25.30   & 0.7623   & 0.2799   \\
IRCNN         &  25.87   & 0.8040  & 0.2436   &  25.62   &  0.8011  & 0.2458   \\
FastDerainNet & 25.76   & 0.7701   & 0.2484   & 21.17   & 0.7585   & 0.2589   \\
DRDNet        & 24.49   & 0.7704   & 0.2650   & 20.61   & 0.7605   & 0.2735   \\
PSLNet         & \textcolor{blue}{27.87}   & \textcolor{blue}{0.8310}   & \textcolor{blue}{0.2105}   & \textcolor{red}{28.03}   & \textcolor{blue}{0.8329}   & \textcolor{blue}{0.2088}   \\
SSH-Net (Ours)     & \textcolor{red}{28.88}  & \textcolor{red}{0.8385} & \textcolor{red}{0.1949} & \textcolor{blue}{27.72}  & \textcolor{red}{0.8336} & \textcolor{red}{0.1993} \\
\hline
\end{tabular}
}
\vspace{-0.15in}
\end{table}

\begin{table}[b]
\caption{PSNR, SSIM, and LPIPS comparison for different methods fixed watermark transparency of 0.3 with a blind noise level of 0, 15, 25 and 50. }
\label{table:3}
\resizebox{1.0\linewidth}{!}{
\begin{tabular}{ccccccc}
\hline
Methods             & PSNR$\uparrow$   & SSIM$\uparrow$   & LPIPS$\downarrow$  &  PSNR$\uparrow$   & SSIM$\uparrow$   & LPIPS$\downarrow$   \\ \hline
Noise levels           & \multicolumn{3}{c}{$\sigma$ = 0}  & \multicolumn{3}{c}{$\sigma$ = 15} \\ \hline
DnCNN         & 35.13  & 0.9794 & \textcolor{blue}{0.0205} &  29.86 & 0.8652 & 0.1648 \\
FFDNet        & 27.39  & 0.8564 & 0.1548 & 26.91  & 0.8048 & 0.2224 \\
IRCNN         & 32.61  & 0.9684 & 0.0335 & 29.10  & 0.8624 & 0.1747 \\
FastDerainNet & 29.85  & 0.9336 & 0.0714 & 27.88  & 0.8352 & 0.1863 \\
DRDNet        & 31.56  & 0.9516 & 0.0517 & 29.40  & 0.8578 & 0.1699 \\
PSLNet        & \textcolor{blue}{35.55}  & \textcolor{blue}{0.9732} & 0.0273 & \textcolor{blue}{30.99}  & \textcolor{blue}{0.8866} & \textcolor{blue}{0.1433} \\ 
SSH-Net (Ours)     & \textcolor{red}{42.59}  & \textcolor{red}{0.9896} & \textcolor{red}{0.0080} & \textcolor{red}{32.18}  & \textcolor{red}{0.8989} & \textcolor{red}{0.1271} \\
\hline
Noise levels           & \multicolumn{3}{c}{$\sigma$ = 25} & \multicolumn{3}{c}{$\sigma$ = 50} \\ \hline
DnCNN         & 27.87  & 0.7951 & 0.2379 & 24.73  & 0.6449 & 0.3645 \\
FFDNet        & 26.27  & 0.7648 & 0.2693 & 24.69  & 0.6778 & 0.3701 \\
IRCNN         & 27.60  &  0.8032 & 0.2416 & 25.28  & 0.6917 & 0.3600 \\
FastDerainNet & 26.70  & 0.7760 & 0.2480 & 24.59  & 0.6474 & 0.3598 \\
DRDNet        & 27.65  & 0.7908 & 0.2375 & 24.63  & 0.6529 &  0.3545 \\
PSLNet         & \textcolor{blue}{29.13}  & \textcolor{blue}{0.8346} & \textcolor{blue}{0.2059} & \textcolor{blue}{26.44}  & \textcolor{blue}{0.7361} & \textcolor{blue}{0.3124} \\ 
SSH-Net (Ours)     & \textcolor{red}{29.95}  & \textcolor{red}{0.8503} & \textcolor{red}{0.1822} & \textcolor{red}{27.10}  & \textcolor{red}{0.7587} & \textcolor{red}{0.2807} \\
\hline
\end{tabular}
}
\vspace{-0.15in}
\end{table}

\label{s:E}
\subsection{Experimental Datasets and Metrics}
We follow the dataset settings in~\cite{tian2024perceptive} and utilize the benchmark datasets for both training and testing. The datasets are constructed using twelve predefined watermarks with varying properties to simulate diverse watermarking scenarios: the training dataset includes 477 natural images from PASCAL VOC 2021~\cite{everingham2015pascal}, each overlaid with one random watermark having transparency levels (0.3, 0.5, 0.7, 1.0), coverage (0 to 0.4), scales (0.5 to 1.0), and Gaussian noise at random levels (0, 15, 25, or 50), while the test dataset comprises 21 images from PASCAL VOC 2012, each processed with a random watermark and a fixed Gaussian noise level, resulting in a total of 252 test images at each test scenario.

For quantitative metrics, following~\cite{tian2024perceptive}, we use peak signal-to-noise ratio (PSNR)~\cite{hore2010image}, structural similarity index (SSIM)~\cite{wang2004image} on the Y channel of the YCbCr color space and learned perceptual image patch similarity (LPIPS)~\cite{zhang2018perceptual} to measure the quality of the results.


\subsection{Experimental Settings}
We set the block number in each level of the shared encoder (SE), in the noise removal decoder (NRD) and in the watermark and noise removal decoder (WNRD) as $(L_1, L_2, L_3, L_4, L_5) = (2,4,4,6,6)$. For the sparse self-attention, we set $K=4$ and the corresponding top-$k$ rate at $(1/2, 2/3, 3/4, 4/5)$. 
The channel number is set to 48 in both the shared encoder and all the decoders, providing a balanced trade-off between computational efficiency and model capacity. In the Sparse Transformer U-Net, the number of attention heads is configured as $(4, 8, 8, 8, 4)$, and the FFN expansion ratio is set to $2.66$.
We set the $\alpha=0.024$ in the mixed loss and train our model using ADAM optimizer with $\beta_{1}=0.9$ and $\beta_{2} = 0.999$. 
The batch size is set to 8, with training conducted over a total of 100 epochs. The initial learning rate is $1\times10^{-3}$, reduced by a factor of 0.1 every 30 epochs to ensure gradual optimization.
All experiments are conducted using PyTorch on a Linux system, with training performed on two NVIDIA RTX A5000 GPUs, while inference is carried out on a single NVIDIA RTX A5000 GPU.


\begin{figure*}[]
\centering
\includegraphics[width=0.97\linewidth]{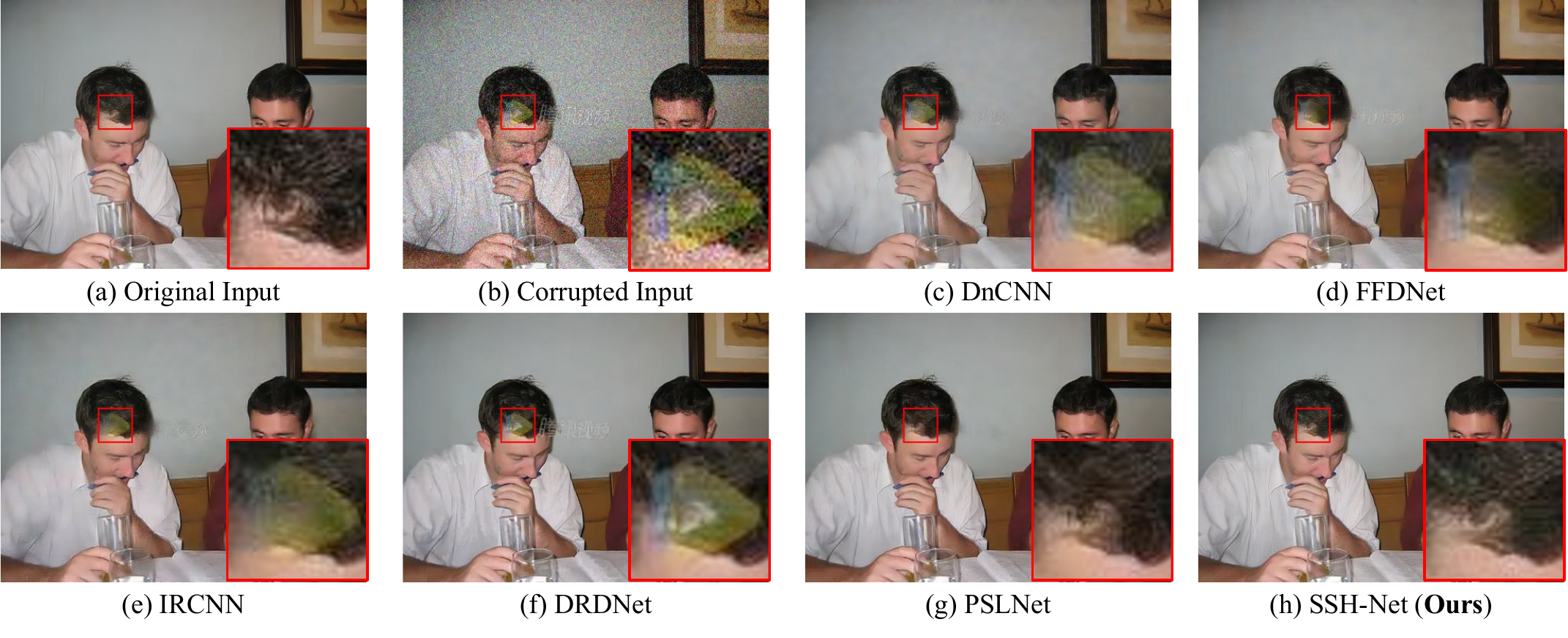}
\vspace{-0.05in}
    \caption{Results comparison trained under a specific transparency and a specific noise condition ($\delta$ = 25 and transparency = 0.3). (a) Ground Truth (b) 20.19 dB (c) 28.48 dB (d) 29.89 dB (e) 29.68 dB (f) 31.39 dB (g) 32.15 dB (h) 32.26 dB. }
    \label{fig:img1}
    \vspace{-0.1in}
\end{figure*}

\begin{figure*}[t]
\centering
\includegraphics[width=0.97\linewidth]{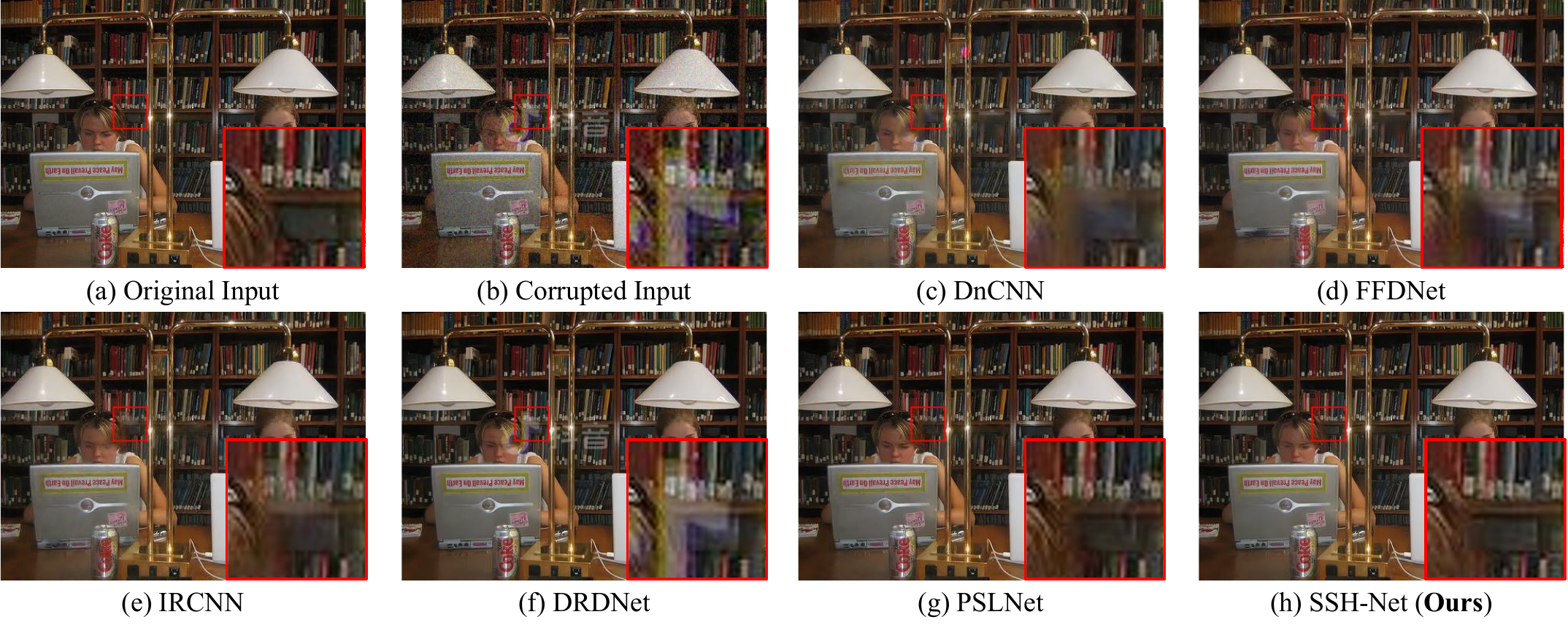}
\vspace{-0.05in}
    \caption{Results comparison trained under a specific noise and a specific transparency condition ($\delta$ = 15 and transparency = 0.3). (a) Ground Truth (b) 24.04 dB (c) 27.44 dB (d) 27.19 dB (e) 30.16 dB (f) 29.23 dB (g) 31.54 dB (h) 31.95 dB. }
    \label{fig:img2}
    \vspace{-0.1in}
\end{figure*}

\subsection{Comparisons with State-of-the-Art Methods}

We evaluate the combined image noise and watermark removal performance of our proposed method through both quantitative and qualitative analyses. To ensure a comprehensive comparison, we benchmark our approach against several widely used image restoration and watermark removal methods, including DnCNN~\cite{zhang2017beyond} and FFDNet~\cite{zhang2018ffdnet} for image denoising, IRCNN~\cite{zhang2017learning} for general image restoration, FastDerainNet~\cite{wang2020fastderainnet} and DRDNet~\cite{deng2019drd} for rain streak removal, and PSLNet~\cite{tian2024perceptive} for combined image noise and watermark removal.

\textbf{\textit{Quantitative Results.}} 
Firstly, we evaluate our model's performance for watermark removal under specific transparency and noise conditions. Specifically, we train and test the model using noise levels of 0, 15, 25, and 50, while keeping the watermark transparency fixed at 0.3. The corresponding results are given in Table~\ref{table:1}.
As shown in the results, the proposed model consistently outperforms the baseline methods in terms of PSNR, SSIM, and LPIPS at all evaluated noise levels, and the performance improvement becomes more obvious at lower noise levels. 
Notably, at a noise level of 0, our approach achieves a remarkable improvement of over 7 dB in PSNR, highlighting its strong capability for watermark removal.
Secondly, we evaluate our model for watermark removal under blind transparency conditions with a fixed noise level. During training, the noise level is set to 25, and the watermark transparency is allowed to vary randomly between 0.3 and 1.0. For testing, we evaluate the model at the same noise level $(25)$ with four fixed transparency values $(0.3, 0.5, 0.7, 1.0)$.
The results, presented in Table~\ref{table:2}, demonstrate that our model consistently achieves robust performance across varying transparency levels in the blind setting, significantly outperforming the compared methods, particularly at transparency levels 0.5 and 0.7.
Thirdly, we evaluate our model for watermark removal under specific transparency with blind noise conditions. During training, we use a fixed watermark transparency of 0.3 and vary the noise levels randomly from 0 to 55. For testing, we evaluate the performance of our model on noisy image removal using fixed noise levels of 0, 15, 25, and 50 with the same watermark transparency of 0.3. The results, presented in Table~\ref{table:3}, illustrate that our model achieves the best results in the watermark removal process, effectively eliminating the watermark across all noise levels. Notably, at lower noise levels (e.g., 0 and 15), our model achieves remarkable improvements in terms of PSNR, demonstrating its superiority in watermark removal with small noise levels.
Fourthly, we evaluate our model for watermark removal under blind transparency with blind noise conditions, which represents the most challenging scenario. During training, the watermark transparency varies randomly between 0.3 and 1.0, while the noise levels are randomly sampled from 0 to 55. For testing, the model is assessed using fixed noise levels of 0, 15, 25, and 50 with a fixed watermark transparency of 0.3, 0.5, 0.7, and 1.0. The results, presented in Table~\ref{tab:4} and Table~\ref{tab:5}, illustrate that even in this most challenging setting, our model significantly outperforms existing methods in all cases, demonstrating its ability to adaptively learn the underlying image content while effectively mitigating the effects of both watermarks and noise. 
In addition, to verify the effectiveness of our method for noise removal without watermarks, we train a single model using varying noise levels from 0 to 55 and evaluate the image denoising performance on fixed noise levels of 15, 25, and 50. The results, presented in Table~\ref{tab:6}, demonstrate that our model outperforms other methods consistently across all tested noise levels.
Lastly, to evaluate the effectiveness of our method for watermark removal without noise, we train a single model using varying watermark transparency levels between 0.3 and 1.0. The model is then tested on fixed transparency levels of 0.3, 0.5, 0.7, and 1.0. The results, presented in Table~\ref{tab:7}, clearly show that our method outperforms existing techniques in effectively eliminating watermarks across nearly all tested transparency levels.

\begin{table}[t]
\caption{PSNR, SSIM, and LPIPS comparison for different methods for blind noise levels and blind watermark transparency.}
\label{tab:4}
\resizebox{1.0\linewidth}{!}{
\begin{tabular}{ccccccc}
\hline
Methods               & PSNR$\uparrow$   & SSIM$\uparrow$   & LPIPS$\downarrow$  &  PSNR$\uparrow$   & SSIM$\uparrow$   & LPIPS$\downarrow$  \\ \hline
Noise levels           & \multicolumn{3}{c}{$\sigma$ = 0}  & \multicolumn{3}{c}{$\sigma$ = 15} \\ \hline
DnCNN         & 28.65  & 0.9590 & 0.0436 & 26.53  & 0.8413 & 0.1855 \\
FFDNet        & 26.04  & 0.8173 & 0.1959 & 25.71  & 0.7829 & 0.2423 \\
IRCNN         & 29.28  & 0.9635 & 0.0386 & 26.94 & 0.8569 & 0.1834 \\
FastDerainNet & 27.24  & 0.9134 & 0.0680 & 26.17  & 0.8174 & 0.1845 \\
DRDNet        & 24.27  & 0.8873 & 0.1108 & 23.10  & 0.7935 & 0.2270 \\
PSLNet         & \textcolor{blue}{35.93}  & \textcolor{blue}{0.9777} & \textcolor{blue}{0.0176} & \textcolor{blue}{31.09}  & \textcolor{blue}{0.8887} & \textcolor{blue}{0.1419} \\ 
SSH-Net (Ours)     & \textcolor{red}{39.42}  & \textcolor{red}{0.9836} & \textcolor{red}{0.0169} & \textcolor{red}{31.81}  & \textcolor{red}{0.8948} & \textcolor{red}{0.1312} \\

\hline
Noise levels           & \multicolumn{3}{c}{$\sigma$ = 25} & \multicolumn{3}{c}{$\sigma$ = 50} \\ \hline
DnCNN         & 25.22  & 0.7624 & 0.2599 & 22.43  & 0.5913 & 0.3955 \\
FFDNet        & 25.25  & 0.7511 & 0.2825 & 24.02 & 0.6781 & 0.3699 \\
IRCNN         & 25.70  & 0.7977 & 0.2528 & 23.45  & 0.6779 & 0.3837 \\
FastDerainNet & 25.23  & 0.7542 & 0.2491 & 23.21  & 0.6143 & 0.3695 \\
DRDNet        & 21.59  & 0.7158 & 0.2987 & 19.02  & 0.5681 & 0.4260 \\
PSLNet      & \textcolor{blue}{29.27}  & \textcolor{blue}{0.8382} & \textcolor{blue}{0.2041} & \textcolor{blue}{26.58}  & \textcolor{blue}{0.7405} & \textcolor{blue}{0.3143} \\ 
SSH-Net (Ours)     & \textcolor{red}{29.72}  & \textcolor{red}{0.8449} & \textcolor{red}{0.1882} & \textcolor{red}{26.92}  & \textcolor{red}{0.7513} & \textcolor{red}{0.2890} \\
\hline
\end{tabular}
}
\vspace{-0.2in}
\end{table}

\begin{table*}[t]
\centering
\caption{PSNR, SSIM, and LPIPS comparison of different methods under blind noise level and blind watermark transparency. Tested at a fixed noise level of 25 and certain watermark transparency of 0.5, 0.7, and 1.0.}
\label{tab:5}
\begin{tabular}{cccccccccc}
\hline
Methods                 & PSNR$\uparrow$   & SSIM$\uparrow$   & LPIPS$\downarrow$  &  PSNR$\uparrow$   & SSIM$\uparrow$   & LPIPS$\downarrow$   & PSNR$\uparrow$   & SSIM$\uparrow$   & LPIPS$\downarrow$    \\ \hline
Transparency          & \multicolumn{3}{c}{Alpha = 0.5} & \multicolumn{3}{c}{Alpha = 0.7} & \multicolumn{3}{c}{Alpha = 1.0} \\ \hline
DnCNN         & 23.88   & 0.7567   & 0.2649   & 23.22   & 0.7555   & 0.2659   & 25.29   & 0.7601   & 0.2622   \\
FFDNet        & 25.27   & 0.7514   & 0.2832   & 25.23   & 0.7501   & 0.2848   & 21.20   & 0.7379   & 0.2944   \\
IRCNN         &  25.52   & 0.7951   & 0.2557   & 25.29   &  0.7927   & 0.2578   & 24.96   & 0.7900   & 0.2599   \\
FastDerainNet & 25.27   & 0.7529   & 0.2505   & 25.07   & 0.7505   & 0.2530   & 20.76   & 0.7397   & 0.2633   \\
DRDNet        & 21.52   & 0.7130   & 0.3026   & 21.77   & 0.7149   & 0.3002   & 18.25   & 0.7064   & 0.3085   \\
PSLNet   & \textcolor{blue}{29.05}   & \textcolor{blue}{0.8364}   & \textcolor{blue}{0.2061}   & \textcolor{blue}{28.61}   & \textcolor{blue}{0.8341}   & \textcolor{blue}{0.2084}   & \textcolor{blue}{27.90}   & \textcolor{blue}{0.8308}   & \textcolor{blue}{0.2116}   \\ 
SSH-Net  (Ours)       & \textcolor{red}{29.62}   & \textcolor{red}{0.8439}   & \textcolor{red}{0.1894}   & \textcolor{red}{29.44}   & \textcolor{red}{0.8422}   & \textcolor{red}{0.1912}   & \textcolor{red}{28.49}   & \textcolor{red}{0.8373}   & \textcolor{red}{0.1963} \\
\hline
\end{tabular}
\vspace{-0.1in}
\end{table*}

\begin{table*}
\centering
\caption{PSNR, SSIM, and LPIPS comparison of different methods under blind noise levels without watermarks. Tested with certain noise levels of 15, 25, and 50 without watermarks.}
\label{tab:6}
\begin{tabular}{cccccccccc}
\hline& PSNR$\uparrow$   & SSIM$\uparrow$   & LPIPS$\downarrow$  &  PSNR$\uparrow$   & SSIM$\uparrow$   & LPIPS$\downarrow$   & PSNR$\uparrow$   & SSIM$\uparrow$   & LPIPS$\downarrow$   \\ \hline
Noise levels           & \multicolumn{3}{c}{$\sigma$ = 15} & \multicolumn{3}{c}{$\sigma$ = 25} & \multicolumn{3}{c}{$\sigma$ = 50} \\ \hline
DnCNN         & 30.42  & 0.8715 & 0.1581 & 28.26  & 0.8011 & 0.2319 & 24.94  & 0.6501 & 0.3599 \\
FFDNet        & 27.50  & 0.8125 & 0.2142 & 26.76  & 0.7716 & 0.2624 & 25.03  & 0.6832 & 0.3655 \\
IRCNN         & 30.78  & 0.8737 & 0.1634 & 28.73  & 0.8131 & 0.2325 & 25.95  & 0.6695 & 0.3543 \\
FastDerainNet & 28.53  & 0.8421 & 0.1792 & 27.24  & 0.7825 & 0.2418 & 24.98  & 0.6531 & 0.3550 \\
DRDNet        & 29.81  & 0.8624 & 0.1654 & 27.92  & 0.7948 & 0.2337 & 24.81  & 0.6564 & 0.3512 \\
PSLNet        & \textcolor{blue}{31.68}  & \textcolor{blue}{0.8919} & \textcolor{blue}{0.1379} & \textcolor{blue}{29.58}  & \textcolor{blue}{0.8395} & \textcolor{blue}{0.2012} & \textcolor{blue}{26.72}  & \textcolor{blue}{0.7406} & \textcolor{blue}{0.3082} \\ 
SSH-Net (Ours)         & \textcolor{red}{32.40}  & \textcolor{red}{0.9012} & \textcolor{red}{0.1241} & \textcolor{red}{30.10}  & \textcolor{red}{0.8526} & \textcolor{red}{0.1794} & \textcolor{red}{27.22}  & \textcolor{red}{0.7612} & \textcolor{red}{0.2780} \\ 
\hline
\end{tabular}
\vspace{-0.1in}
\end{table*}

\begin{table}[t]
\caption{PSNR, SSIM, and LPIPS of different methods trained blind watermark transparency without noise. Tested with certain watermark transparency of 0.3, 0.5. 0.7, 1.0 without noise.}
\label{tab:7}
\resizebox{1.0\linewidth}{!}{
\begin{tabular}{ccccccc}
\hline
Methods               & PSNR$\uparrow$   & SSIM$\uparrow$   & LPIPS$\downarrow$  &  PSNR$\uparrow$   & SSIM$\uparrow$   & LPIPS$\downarrow$      \\ \hline
Transparency          & \multicolumn{3}{c}{Alpha = 0.3} & \multicolumn{3}{c}{Alpha = 0.5} \\ \hline
DnCNN         & 29.49   & 0.9406   & 0.0617   & 29.39   & 0.9405   & 0.0614   \\
FFDNet        & 25.87   & 0.8548   & 0.1410   & 25.83   & 0.8551   & 0.1408   \\
IRCNN        & 31.21   & 0.9673   & 0.0264   & 31.12   & 0.9659   & 0.0279   \\
FastDerainNet & 26.97   & 0.9508   & 0.0274   & 26.58   & 0.9504   & 0.0263   \\
DRDNet        & 31.02   & 0.9763   & 0.0267   & 31.26   & 0.9752   & 0.0284   \\
PSLNet       & \textcolor{blue}{38.66}   & \textcolor{blue}{0.9909}   & \textcolor{blue}{0.0075}   & \textcolor{blue}{38.48}   & \textcolor{blue}{0.9903}   & \textcolor{blue}{0.0081}   \\ 
SSH-Net (Ours)       & \textcolor{red}{43.19}   & \textcolor{red}{0.9961}   & \textcolor{red}{0.0043}   & \textcolor{red}{43.04}   & \textcolor{red}{0.9959}   & \textcolor{red}{0.0047}   \\ 

\hline
Transparency          & \multicolumn{3}{c}{Alpha = 0.7} & \multicolumn{3}{c}{Alpha = 1.0} \\ \hline
DnCNN         & 29.21   & 0.9389   & 0.0633   & 22.22   & 0.9211   & 0.0819   \\
FFDNet        & 25.83   & 0.8536   & 0.1425   & 21.43   & 0.8406   & 0.1546   \\
IRCNN         & 30.95  & 0.9637   & 0.0306   & 29.71  & 0.9585   & 0.0361   \\
FastDerainNet & 26.11   & 0.9479   & 0.0285   & 21.27   & 0.9358   & 0.0433   \\
DRDNet        & 29.92   & 0.9709   & 0.0330   & 22.91   & 0.9605   & 0.0430   \\
PSLNet        & \textcolor{blue}{37.40}   & \textcolor{blue}{0.9884}   & \textcolor{blue}{0.0103}   & \textcolor{red}{34.50}   & \textcolor{blue}{0.9820}   & \textcolor{blue}{0.0177}   \\ 
SSH-Net (Ours)   & \textcolor{red}{41.73}   & \textcolor{red}{0.9944}   & \textcolor{red}{0.0063}   & \textcolor{blue}{33.85}   & \textcolor{red}{0.9842}   & \textcolor{red}{0.0175}   \\ 

\hline
\end{tabular}
}
\vspace{-0.1in}
\end{table}

\begin{table}[]
\caption{Computational complexity comparison of different methods on a 256$\times$256 image.}
\label{tab:8}
\resizebox{0.8\linewidth}{!}{
\begin{tabular}{ccc}
\hline
Methods   & \#Parameters    & \#FLOPs   \\ 
\hline 
DnCNN  & 0.56M & 36.59G \\
DRDNet & 2.94M & 192.49G \\
PSLNet & 2.52M & 74.51G \\
SSH-Net (Ours) & 5.89M & 18.21G \\
\hline 
\end{tabular}
}
\vspace{-0.15in}
\end{table}


\textbf{\textit{Qualitative Results.}} 
In Fig.~\ref{fig:img1} and Fig.~\ref{fig:img2}, we qualitatively compare our model with baseline methods, including DnCNN, FFDNet, IRCNN, DRDNet, and PSLNet. Fig.~\ref{fig:img1} illustrates the restored images at a specific noise level of 25 and a specific transparency level of 0.5. As observed, while most methods, such as DnCNN and DRDNet, effectively reduce noise, they often struggle with watermark removal. In contrast, both PSLNet and our proposed model successfully eliminate noise and watermarks simultaneously while preserving the structural integrity and fine details of the images. Moreover, compared to PSLNet, our method delivers superior visual fidelity and demonstrates enhanced detail preservation across various regions of the images. 
Fig.~\ref{fig:img2} shows the restored images at a specific noise level of 15 and a specific transparency level of 0.5. As the noise level decreases, most methods show improved performance, with more baseline methods capable of removing both noise and watermarks simultaneously. But, they continue to struggle with preserving fine textural details and maintaining the overall coherence of the image. In contrast, our proposed model delivers visually superior results, effectively achieving both detail preservation and comprehensive noise and watermark removal.


\begin{table}[b]
\caption{Runtime and GPU memory consumption of different methods evaluated on 256$\times$256 images.}
\label{tab:12}
\resizebox{0.85\linewidth}{!}{
\begin{tabular}{ccc}
\hline
Methods   & \#Runtime    & \#GPU Memory   \\ 
\hline 
DnCNN  & 2.03ms & 18.78M \\
DRDNet & 36.78ms & 183.06M \\
PSLNet & 10.45ms & 330.72M \\
SSH-Net (Ours) & 43.03ms & 197.46M \\
\hline 
\end{tabular}
}
\vspace{-0.15in}
\end{table}



\begin{figure}[]
\centering
\includegraphics[width=0.98\linewidth]{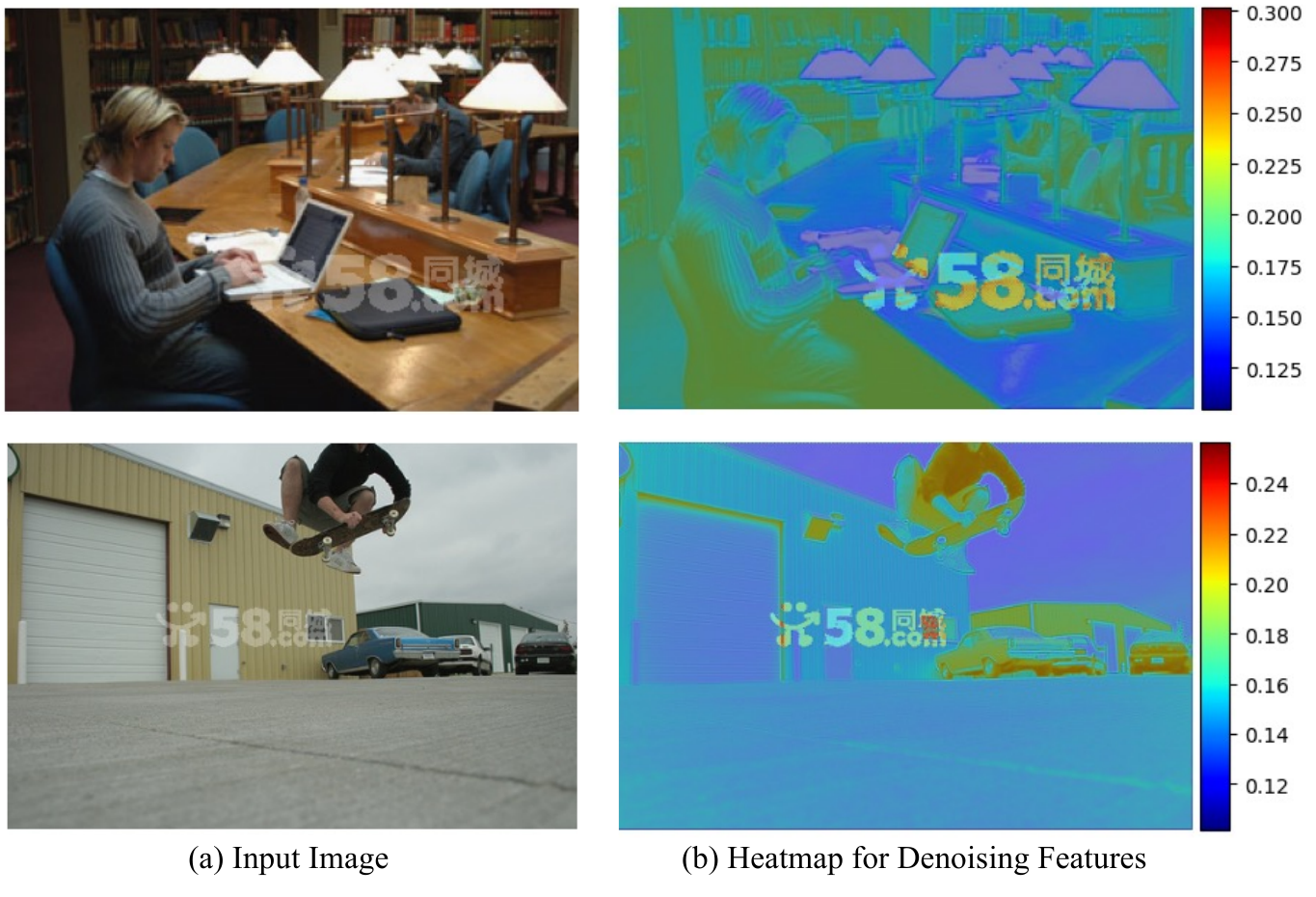}
\vspace{-0.05in}
\caption{Visualizations of the gating mechanism in the proposed Feature Fusion Unit (FFU).}
    \vspace{-0.2in}
    \label{fig:img11}
\end{figure}

\textbf{\textit{Complexity Comparison.}} 
To verify the practicality of our SSH-Net for digital devices, we evaluate its parameters and computational efficiency by analyzing the number of parameters and FLOPs, and compare it with baseline methods, including DnCNN~\cite{zhang2017beyond}, DRDNet~\cite{deng2019drd} and PSLNet~\cite{tian2024perceptive}.
As shown in Table~\ref{tab:8}, although our SSH-Net has a larger number of parameters than DnCNN, DRDNet, and PSLNet, it achieves a stronger representational capacity, delivering significantly better watermark removal performance across various noise and transparency levels. Furthermore, SSH-Net achieves fewer FLOPs than DnCNN, DRDNet, and PSLNet, owing to its hybrid structure design and the incorporation of the proposed Sparse Transformer U-Net, ensuring computational efficiency.
We also evaluate the runtime and GPU memory consumption for different models conducted on a single NVIDIA GeForce RTX A5000 and compare the performance metrics outlined in Table~\ref{tab:12}. Our experimental results indicate that although SSH-Net exhibits lower FLOPs compared to other methods, its runtime is slightly higher. This is mainly due to the use of non-convolutional operations (e.g., sparse attention) and a dual-branch structure, in which the two decoders have different computational complexities. This imbalance leads to suboptimal parallelism during inference, which, along with increased memory access from the shared encoder, affects practical execution speed despite the lower theoretical computation.

\textbf{\textit{Visualization of Gating Signal.}} 
To better understand the behavior of the gating mechanism, we visualize the spatial activation of the gating signal inside the proposed Feature Fusion Unit (FFU) in the Fig.~\ref{fig:img11}. This gating signal is learned from the watermark removal pathway and is applied to modulate the denoising features $F_\text{n}$ before fusion $F_\text{n}$. As shown in the Fig.~\ref{fig:img11}, the gating map is obtained by averaging the gating weights across all channels, resulting in a spatially varying response across different regions of the image. This behavior suggests that the model has learned to adaptively adjust the contribution of denoising features based on local content quality. For instance, in regions where the denoising pathway performs well, the gate assigns higher weights, allowing more information from the denoising branch to pass through. Conversely, in areas where denoising results are less reliable, the gate assigns lower weights, relying more on the complementary features from the watermark branch.
More importantly, the majority of the gating weights are significantly smaller than 0.5, indicating that the $F_\text{n}$ are only partially preserved, serving as a complementary component rather than a dominant one during integration.

\subsection{Ablation Study}
In this section, we present an ablation study to analyze the individual contributions of the components in the SSH-Net architecture. We systematically remove or modify each module and assess the impact on the overall performance metrics, including PSNR and SSIM.

\textbf{\textit{Effect of Proposed Components.}} 
We evaluate the effectiveness of different components by constructing several variants of SSH-Net with specific loss settings: (a) SE with NRD, generating one output and optimized with structure and texture loss relative to $Y_\text{w}$; (b) SE with WNRD, generating one output and optimized with structure and texture loss relative to $Y_\text{w}$; (c) SE with both NRD and WNRD, generating two outputs and optimized with structure and texture loss relative to $(X_\text{w}, Y_\text{w})$; and (d) the full SSH-Net, optimized with the complete loss formulation for all three outputs. 
We train and test all these models under a specific noise and a specific transparency condition ($\delta$ = 25 and transparency = 0.3).
As shown in Table~\ref{tab:9}, the model with WNRD demonstrates significantly higher effectiveness compared to NRD, attributed to the application of Transformers, which enhance the modeling of long-range dependencies. Moreover, the full SSH-Net achieves the best performance across all metrics, highlighting the advantages of integrating NRD, WNRD, and the FFU together.

\begin{table}[]
\caption{Ablation Study on individual components in the SSH-Net architecture under a specific noise and a specific transparency condition ($\delta$ = 25 and transparency = 0.3).}
\label{tab:9}
\resizebox{1.0\linewidth}{!}{
\begin{tabular}{ccccccc}
\hline
Models   & SE & NRD & WNRD & FFU  & PSNR $\uparrow$ / SSIM $\uparrow$  \\ 
\hline 
(a) & \cmark & \cmark & \xmark & \xmark & 29.15 / 0.8327 \\
(b) & \cmark & \xmark & \cmark & \xmark & 29.73 / 0.8484 \\
(c) & \cmark & \cmark & \cmark & \xmark & 29.83 / 0.8530 \\
(d) & \cmark & \cmark & \cmark & \cmark & 29.86 / 0.8533   \\
\hline 
\end{tabular}
}
\vspace{-0.1in}
\end{table}

\textbf{\textit{Effect of Shared Encoder.}} We evaluate the effect of using a shared encoder (SE) by comparing it with a dual-encoder setup under the same training and evaluation conditions. The results, presented in Table~\ref{tab:11}, show that using the SE significantly reduces both the number of parameters and the computational cost. Since the encoder processes the full-resolution input image from the beginning, duplicating it for two branches would result in substantial redundant computation. In contrast, the shared encoder enables efficient feature reuse across both decoders without sacrificing performance. Moreover, the SE configuration slightly outperforms the dual-encoder counterpart in PSNR and SSIM, likely due to better feature alignment and consistency enabled by the unified representation.


\begin{table}[]
\caption{Ablation study comparing shared encoder and dual encoder configurations. Experiments are conducted on 256$\times$256 images under a specific noise and a specific transparency condition ($\delta$ = 25 and transparency = 0.3).}
\label{tab:11}
\resizebox{\linewidth}{!}{
\begin{tabular}{cccc}
\hline
Methods   & \#Parameters    & \#FLOPs     & PSNR $\uparrow$ / SSIM $\uparrow$ \\ 
\hline 
Dual encoders  & 6.04M & 21.62G & 29.80 / 0.8529\\
Ours & 5.89M & 18.21G  & 29.86 / 0.8533 \\
\hline 
\end{tabular}
}
\vspace{-0.15in}
\end{table}

\begin{table}[]
\caption{Ablation study on the effect of Sparse Self-Attention, trained under blind noise and blind transparency conditions, and tested at a fixed noise level of 25 with watermark transparency levels of 0.5, 0.7, and 1.0.}
\label{tab:10}
\resizebox{1.0\linewidth}{!}{
\begin{tabular}{cccc}
\hline
Models   & Alpha = 0.5 & Alpha = 0.7 & Alpha = 1.0  \\ 
\hline 
w/o SSA & 28.98 / 0.8372  & 28.80 / 0.8353 & 28.04 / 0.8305\\
Ours & 29.62 / 0.8439 & 29.44 / 0.8422 & 28.49 / 0.8373 \\ \hline 
\end{tabular}
}
\vspace{-0.15in}
\end{table}

\textbf{\textit{Effect of Sparse Self-Attention.}} 
We also evaluate the effect of the sparse self-attention mechanism, as shown in Table~\ref{tab:10}. To this end, we construct a baseline model by replacing the sparse self-attention layer with a multi-Dconv head transposed attention (MDTA) layer~\cite{zamir2022restormer}. Both the baseline model and the full SSH-Net are trained under blind noise and blind transparency conditions and tested at a fixed noise level of 25 with varying watermark transparency levels. 
The results show that introducing sparse self-attention into the proposed Sparse Transformer U-Net significantly enhances performance, compared to Standard MDTA, demonstrating its effectiveness in both feature representation and modeling for image noise and watermark removal.

\section{Conclusion}
\label{s:C}
In this paper, we propose SSH-Net, a Self-Supervised and Hybrid Network for noisy image watermark removal. Unlike existing methods that rely on paired watermarked and watermark-free images, SSH-Net synthesizes reference watermark-free images in a self-supervised manner based on the watermark distribution. The architecture features a shared encoder for shallow feature extraction, a noise removal decoder to handle noise exclusively, a watermark and noise removal decoder to address both watermarks and noise, and a feature fusion unit to integrate feature maps from both decoders. To enhance performance, the sparse Transformer U-Net is incorporated into the watermark and noise removal decoder, leveraging sparse self-attention to effectively capture long-range dependencies and focus on relevant features while suppressing irrelevant ones. 
This design allows SSH-Net to achieve superior performance in noisy image watermark removal, as demonstrated by experimental results that highlight its advantage over state-of-the-art methods in both visual quality and quantitative metrics, while maintaining a compact computation cost.



{
\bibliographystyle{plainnat}
\bibliography{cas-refs}
}

\end{document}